\documentclass[12pt,a4paper]{article}
\usepackage[truedimen,margin=30mm]{geometry} 
\usepackage[utf8]{inputenc}

\usepackage{float}
\usepackage{amsmath}
\usepackage{amsthm}
\usepackage{amssymb}
\usepackage{setspace}
\usepackage[authoryear]{natbib}
\usepackage{graphicx}
\usepackage{color}
\usepackage{mathrsfs}
\usepackage{bm}
\usepackage{url}
\usepackage{booktabs}
\usepackage{multirow}
\usepackage{multicol}
\usepackage{hhline}
\usepackage{appendix}

\usepackage{titlesec}
\titleformat*{\section}{\large\bfseries}
\titleformat*{\subsection}{\it}

\providecommand{\remarkname}{Remark}
\theoremstyle{remark}
\newtheorem{rem}{\protect\remarkname}
\title{{\bf Bayesian spatiotemporal conditional autoregressive model for local temporal variations}}

\author{Takahiro Onizuka$^1$ and Shintaro Hashimoto$^2$\\
$^1$Graduate School of Social Sciences, Chiba University, Japan\\
$^2$Department of Mathematics, Hiroshima University, Japan}

\date{}

\begin{document}

\maketitle


\begin{abstract}
Spatiotemporal areal data are commonly observed in various fields such including epidemiology, social science, economics and so on. To capture both spatial trends and temporal trends, spatiotemporal modeling is often employed, and the conditional autoregressive (CAR) model is one of the most widely used approaches for modeling areal data. This paper proposes a new framework for estimating spatiotemporal trends based on the CAR model. The proposed method provides locally adaptive temporal smoothing while yielding interpretable temporal trends by effectively utilizing information from both spatially neighboring areas and temporally adjacent time points. We also develop a Gibbs sampling algorithm and demonstrate the ability of the proposed method to adapt to to local temporal changes through numerical examples. 
\end{abstract}

\vspace{-0cm}


\section{Introduction}

In recent years, areal spatiotemporal data, observed repeatedly over a set of non-overlapping geographical regions, have become increasingly available in many fields, including epidemiology, environmental science, economics, and the social sciences. Such data often exhibit both spatial dependence among neighboring regions and temporal dependence across adjacent time points. Therefore, statistical models that can simultaneously account for spatial and temporal dependence are essential for reliable inference. Bayesian hierarchical models provide a flexible framework for this purpose, as they can naturally combine an observation model with latent spatiotemporal effects. In particular, conditional autoregressive (CAR) priors have been widely used to model spatial dependence among areal units and have become a standard tool in spatial statistics and disease mapping. CAR priors have also been extensively studied in small area estimation \citep[see, e.g.,][]{fay1979estimates}. \cite{besag1991bayesian} proposed a Bayesian disease mapping model that combines spatially structured and unstructured random effects, thereby establishing the CAR prior as a fundamental component of small-area risk estimation. The Leroux CAR prior proposed by \cite{leroux2000statistical} is also widely used because it provides a flexible transition between spatial independence and strong spatial smoothing through a single spatial dependence parameter. In many spatiotemporal CAR models, the spatial precision matrix is constructed from an adjacency matrix together with a parameter that controls the strength of spatial dependence \citep[see, e.g.,][]{leroux2000statistical}.

Early developments in spatiotemporal CAR modeling include \cite{bernardinelli1995bayesian}, who proposed a model with region-specific linear temporal trends in which spatial dependence is introduced into both the intercepts and slopes. Knorr-Held (2000) further developed a hierarchical framework that decomposes spatiotemporal disease risk into spatial main effects, temporal main effects, and space-time interaction effects, and introduced several prior specifications for the interaction component. These models are useful for decomposing spatiotemporal variation into interpretable components. However, because temporal variation is typically represented by linear trends or additive main and interaction effects, such models may have limited flexibility in capturing abrupt changes or locally nonlinear temporal patterns.

Autoregressive spatiotemporal CAR models provide a more direct approach to modeling temporal dependence in latent spatial surfaces. For example, in the AR-type model of \cite{rushworth2014spatio}, the spatial random effect at each time point is modeled as a multivariate autoregressive process that depends on the spatial random effect at the previous time point, while spatial dependence is represented through a Leroux-type CAR precision matrix. \cite{lee2016quantifying} and \cite{mozdzen2022spatial} also proposed AR-type priors in spatiotemporal CAR models. However, standard AR-type spatiotemporal CAR models impose a globally smooth temporal evolution through a small number of dependence parameters. In particular, temporal smoothing is controlled mainly by a single autoregressive parameter, and spatial smoothing is controlled by a single spatial dependence parameter. As a result, the same degree of temporal and spatial smoothness is imposed over the entire study region and throughout the observation period. This global smoothing structure can be too restrictive when the underlying process exhibits abrupt changes or piecewise polynomial temporal trends. Such features frequently arise in applications involving economic crises, infectious disease outbreaks, policy interventions, and natural disasters, where substantial level shifts may occur at particular time points, either globally or locally across regions. \cite{rushworth2017adaptive} further proposed a spatially adaptive spatiotemporal smoothing model for disease risk estimation in which the smoothing structure between neighboring areas is estimated from the data. \cite{mozdzen2022spatial} introduced spatially clustered temporal correlation parameters into an AR-type spatiotemporal CAR prior.

In this paper, we focus on spatiotemporal smoothing and denoising without explanatory variables, with the aim of estimating latent spatiotemporal effects or underlying trends for individual areas. Spatial smoothing can provide interpretable and stable estimates by borrowing information from neighboring areas, while temporal smoothing methods, such as spline-based approaches, can effectively denoise observations and recover nonlinear underlying trends. Our objective is therefore to estimate interpretable underlying temporal trends for each area while simultaneously exploiting spatial dependence among neighboring areas. To this end, we propose a locally adaptive alternative to conventional AR-type spatiotemporal CAR models by introducing global-local shrinkage priors on temporal differences of latent spatial effects. Specifically, for the latent spatiotemporal effects, we consider shrinkage priors on either first-order or second-order temporal differences. The proposed prior combines a spatial CAR precision matrix with time-specific local scale parameters, thereby allowing the degree of temporal shrinkage to vary over time. When the latent spatial surface evolves smoothly, the temporal differences are strongly shrunk toward zero; when an abrupt level shift or a change in trend occurs, the shrinkage is weakened, allowing the corresponding temporal difference to remain large. Therefore, the proposed model retains the spatial dependence structure of conventional spatiotemporal CAR models while providing additional flexibility to capture piecewise constant or locally adaptive temporal trends. This approach provides stable estimates by borrowing strength across spatially neighboring areas and temporally adjacent time points. Note that the proposed prior can also be applied to small area estimation as mentioned in Section~\ref{sec:concluding}. Global-local shrinkage for the smoothing of time-series data has been considered by \cite{faulkner2018locally}. In the context of small area estimation, \cite{tang2023global} considered area-specific global-local shrinkage, and multivariate extensions have also been proposed \citep[e.g.,][]{ghosh2022multivariate,nishina2026global}. Although \cite{wakayama2023trend} proposed a locally adaptive frequentist smoothing method on graphs, their approach does not consider CAR models and assumes smooth temporal variation within each area and is therefore not designed to accommodate discontinuous temporal trajectories, which are the primary focus of this paper. 

The remainder of the paper is organized as follows. In Section 2, we describe the proposed locally adaptive spatiotemporal CAR models based on global-local shrinkage priors. Section 3 presents simulation studies evaluating the performance of the proposed models. In Section 4, we apply the proposed models to real data examples. Section 5 concludes the paper.

\section{Methodology}

\subsection{Conditional autoregressive (CAR) model}

This paper focuses on spatiotemporal denoising or smoothing without explanatory variables. To this end, let $y_{it}$ denote the observation for area $i$ at time $t$, and consider the following model: 
\begin{equation}
y_{it}=\theta_{it}+\epsilon_{it},\quad \epsilon_{it}\sim N(0, \sigma^2),\quad i=1,\dots,n,\ t=1,\dots,T
\label{obs-eq}
\end{equation}
where $\theta_{it}$ represents the underlying spatial trend in area $i$  at time $t$, and $\epsilon_{it}$ is an i.i.d. error term. For point-referenced data, spatial dependence in $\theta_{it}$ is often modeled using a Gaussian process. In contrast, this paper focuses on areal data, for which the spatial structure is represented by a graph. Specifically, the underlying spatiotemporal trend is assumed to exhibit spatial dependence according to the neighborhood structure determined by whether pairs of areas share a common border. To detect an areal trend, conditional autoregressive (CAR) models are widely used to model spatial dependence. In particular, the Leroux CAR model \citep{leroux2000statistical} specifies the conditional prior distribution 
of $\theta_{it}$ as
\begin{equation*}
    \theta_{it}\mid \theta_{(-i)t}\sim N\left(\frac{\rho\sum_{\ell\neq i}w_{i\ell}\theta_{\ell t}}{\rho\sum_{\ell\neq i} w_{i\ell}+1-\rho}, \frac{\tau^2}{\rho\sum_{\ell\neq i} w_{i\ell}+1-\rho}\right),
\end{equation*}
where $\theta_{(-i)t}$ denotes the $(n-1)$-dimensional vector obtained 
by removing $\theta_{it}$ from $\theta_t=(\theta_{1t},\dots,\theta_{nt})^\top$, and $\rho$ is a spatial  dependence parameter. The neighborhood indicator $w_{ij}$ is equal to one if areas $i$ and $j$ share a common border and zero otherwise. Let $W=(w_{ij})$ denote the $n\times n$ adjacency matrix, and define 
$$
    D=\operatorname{diag}(w_{1+},\dots,w_{n+}),
    \quad w_{i+}=\sum_{j=1}^n w_{ij}.
$$
Using Brook's lemma, the corresponding joint prior distribution of $\theta_{t}=(\theta_{1t},\dots,\theta_{nt})^\top$ is given by
\begin{equation*}
    \theta_t\sim N_n\left(0, \tau^2Q_{(W,\rho)}^{-1}\right),
\end{equation*}
where $Q_{(W,\rho)}=(1-\rho)I_n+\rho (D-W)$, $W$ is an $n\times n$ adjacency matrix, and $I_n$ is an identity matrix. When $\rho=0$, the components of $\theta_t$ are a priori independent, with
$$
    \theta_t\sim N_n(0,\tau^2 I_n).
$$
As $\rho$ increases, greater weight is placed on the spatial neighborhood structure. At $\rho=1$, the precision matrix reduces to the graph Laplacian $D-W$, corresponding to the intrinsic CAR model \citep{besag1991bayesian}. Since $D-W$ is singular, the resulting prior is improper unless an appropriate identifiability constraint, such as a  sum-to-zero constraint, is imposed. Thus, for the proper Leroux CAR  prior, $\rho\in[0,1)$ controls the strength of spatial smoothing.

To incorporate temporal dependence, \cite{rushworth2014spatio} considered  an AR(1)-type extension of the CAR model given by
\begin{equation}
    \theta_1\sim N_n\left(0, \tau^2Q_{(W,\rho)}^{-1}\right),\quad\theta_t\sim N_n\left(\rho_T\theta_{t-1}, \tau^2Q_{(W,\rho)}^{-1}\right),\label{AR1-model}
\end{equation}
where $0\leq\rho_T\leq1$ is a temporal autoregressive parameter. When  $\rho_T=0$, the latent spatial effects are independent across time  conditional on the model parameters, while retaining the same spatial  dependence structure at each time point. This formulation can be viewed  as a multivariate Gaussian Markov random field evolving over time \citep{rue2005gaussian}.

A second-order autoregressive extension can similarly be defined as
\begin{align*}
&\theta_1\sim N_n\left(0, \tau^2Q_{(W,\rho)}^{-1}\right),\quad \theta_2\sim N_n\left(0, \tau^2Q_{(W,\rho)}^{-1}\right),\\
&\theta_t\mid \theta_{t-1},\theta_{t-2}\sim N_n\left(\rho_{T_1}\theta_{t-1}+\rho_{T_2}\theta_{t-2}, \tau^2Q_{(W,\rho)}^{-1}\right),
\end{align*}
where $\rho_{T_1}$ and $\rho_{T_2}$ are temporal autoregressive  parameters. Such models borrow information not only from neighboring  areas but also from temporally adjacent observations, and can therefore  improve estimation accuracy when the underlying spatiotemporal surface  evolves smoothly over time.

\subsection{Hierarchical model for local adaptivity}
From the perspective of global-local shrinkage, the parameter $\tau$ in \eqref{AR1-model} can be regarded as a global scale parameter for the temporal innovations. In addition, $\rho_T$ is a global autoregressive parameter that controls temporal persistence throughout the entire observation period. Therefore, such a global specification may lead to excessive smoothing when the underlying temporal process exhibits local or abrupt changes. Motivated by global-local shrinkage, we propose a first-order locally adaptive spatiotemporal Bayesian CAR model by specifying the following prior distributions for $\theta_t$:
\begin{equation}
    \eta_1=\theta_1\sim N_n\left(0, u_1^2Q_{(W,\rho_1)}^{-1}\right),\quad \eta_t=\theta_t-\theta_{t-1}\sim N_n\left(0, \tau^2u_t^2Q_{(W,\rho_t)}^{-1}\right).
    \label{1st-order}
\end{equation}
Let $\eta=(\eta_1^\top,\dots,\eta_T^\top)^\top\in\mathbb{R}^{nT}$ denote the transformation of $\theta=(\theta_1^\top,\dots,\theta_T^\top)^\top\in\mathbb{R}^{nT}$.
The joint prior distribution of $\eta$ is then given by
$$
p(\eta\mid \tau,u,\rho)=\prod_{t=1}^T(2\pi)^{-n/2}|R_t|^{1/2}\exp\left(-\frac{1}{2}\eta_t^\top R_t\eta_t\right),
$$
where $R_1=u_1^{-2}Q_{(W,\rho_1)}$ and $R_t=(\tau^2u_t^2)^{-1}Q_{(W,\rho_t)}$ for $t\ge 2$. Here, $\tau$ is a global scale parameter, $u_t$ is a time-specific
local scale parameter, and $\rho_t$ is a time-specific spatial dependence parameter. In particular, $u_t$ controls the degree of shrinkage of the temporal difference $\theta_t-\theta_{t-1}$, whereas $\rho_t$ controls the spatial dependence of the temporal changes at time $t$. 
To accommodate local temporal changes, we assign the half-Cauchy prior $u_t\sim C^+(0,1)$ for $t\ge 2$. Combined with the Gaussian scale-mixture representation in \eqref{1st-order}, this specification yields horseshoe-type shrinkage \citep{carvalho2010horseshoe}, which has also been used for locally adaptive smoothing of time-series data \citep[see, e.g.,][]{faulkner2018locally}.
For the remaining parameters, we assume $u_1^2\sim \mathrm{IG}(a_{u_1},b_{u_1})$, $\tau\sim C^+(0,1)$,  and $\sigma^2\sim\mathrm{IG}(a_{\sigma^2},b_{\sigma^2})$, where $\mathrm{IG}(a,b)$ denotes an inverse gamma distribution with shape $a$ and rate $b$. For simplicity, we do not impose an additional temporal dependence
structure on $\rho_t$. Instead, the $\rho_t$'s are assumed to be independent a priori and to follow a discrete uniform distribution on a prespecified grid $[0,1)$.

The proposed framework can also be extended to second-order temporal differences. Specifically, the second-order global-local spatiotemporal Bayesian CAR model is defined by
\begin{align*}
&\eta_1=\theta_1\sim N_n\left(0, u_1^2Q_{(W,\rho_1)}^{-1}\right),\quad \eta_2=\theta_2-\theta_{1}\sim N_n\left(0, \tau^2u_2^2Q_{(W,\rho_2)}^{-1}\right),\\
&\eta_t=\theta_t-2\theta_{t-1}+\theta_{t-2}\sim N_n\left(0, \tau^2u_t^2Q_{(W,\rho_t)}^{-1}\right).
\label{2nd-order}
\end{align*}
Although we focus primarily on the first-order model in this paper, the second-order model can be developed analogously. These formulations can be viewed as multivariate spatial extensions of locally adaptive trend-filtering-type models, in which shrinkage is imposed on temporal differences \citep[e.g.,][]{kim2009ell_1,faulkner2018locally,
onizuka2024bayesian,onizuka2024fast}.

\begin{rem}\label{rem-shrinkage-factor}
For $t\geq2$, after integrating out the local scale parameter $u_t$, the conditional marginal prior of $\theta_t$ given $\theta_{t-1}$, $\tau$, and $\rho_t$ has an unbounded spike at $\theta_t=\theta_{t-1}$ under the horseshoe-type prior. This property encourages strong shrinkage of small temporal differences while allowing large differences to escape shrinkage. The role of $u_t$ can also be understood through the posterior mean.
Let $y_t=(y_{1t},\dots,y_{nt})^\top$ and $Q_t=Q_{(W,\rho_t)}$. Conditional on $\theta_{t-1}$, $u_t$, $\tau$, $\sigma^2$, and $\rho_t$,
the posterior distribution of $\theta_t$ is Gaussian, with mean
\begin{align*}
    E[\theta_t \mid \theta_{t-1},u_t,\tau,\sigma^2,\rho_t,y_t]
    &=\left(I_n+\frac{\sigma^2}{\tau^2u_t^2}Q_t \right)^{-1}
    \left(y_t+\frac{\sigma^2}{\tau^2u_t^2}Q_t\theta_{t-1}\right).
\end{align*}
Following the multivariate shrinkage-factor representation, define
\begin{equation*}
    S_t=\left(I_n+\frac{\sigma^2}{\tau^2u_t^2}Q_t\right)^{-1}\frac{\sigma^2}{\tau^2u_t^2}Q_t.
\end{equation*}
Then the posterior mean can be expressed as
\begin{equation*}
    E[\theta_t\mid \theta_{t-1},u_t,\tau,\sigma^2,\rho_t,y_t]=(I_n-S_t)y_t+S_t\theta_{t-1}.
\end{equation*}
Thus, $S_t$ can be interpreted as a matrix-valued shrinkage factor toward the previous latent spatial trend $\theta_{t-1}$. Consequently, when $u_t$ is large, $S_t$ becomes small and $\theta_t$ is estimated closer to $y_t$; when $u_t$ is small, $S_t$ approaches $I_n$ and $\theta_t$ is strongly shrunk toward $\theta_{t-1}$. See also \cite{tang2023global}. 
\end{rem}

\subsection{Sampling procedure}\label{computation}
The proposed model can be implemented using a straightforward Markov chain Monte Carlo (MCMC) algorithm, specifically a Gibbs sampling algorithm. To facilitate Gibbs sampling, we employ the inverse-gamma mixture representation of the half-Cauchy distribution. Specifically, the priors for the local and global scale parameters can be represented hierarchically as $u_t^2\mid\nu_t\sim \mathrm{IG}(1/2, 1/\nu_t)$, $\nu_t\sim\mathrm{IG}(1/2,1)$, $\tau^2\mid\phi\sim \mathrm{IG}(1/2,1/\phi)$ and $\phi\sim\mathrm{IG}(1/2,1)$. 
Under this representation, the full conditional distributions for the proposed first-order model are given as follows:
\begin{itemize}
\item[--] Sample $\eta_t$ ($t=1,\dots,T$) from $N(\Omega_t^{-1}\mu_t,\Omega_t^{-1})$ with
\begin{align*}
&\Omega_t=(T-t+1)\sigma^{-2}I+R_{t},\quad \mu_t=\sigma^{-2}\sum_{s=t}^T\left(y_s-\sum_{r=1,r\neq t}^s\eta_r\right)
\end{align*}    
\item[--] Sample $\tau^2$ from
$$\tau^2\sim \mathrm{IG}\left(\frac{n(T-1)+1}{2},\sum_{t=2}^T\frac{
\eta_t^{\top}Q_t\eta_t}{2u_t^2}+\frac{1}{\phi}\right)
$$
\item[--] Sample $u_t^2$ from
\begin{align*}
u_1^2&\sim\mathrm{IG}\left(\frac{n}{2}+a_{u_1},
\frac{\eta_1^{\top}Q_1\eta_1}{2} + b_{u_1}\right),\quad
u_t^2\sim\mathrm{IG}\left(\frac{n+1}{2},\frac{
\eta_t^{\top}Q_t\eta_t}{2\tau^2} + \frac{1}{\nu_t}\right)\quad t\ge 2.
\end{align*}
\item[--] Sample $\phi$ and $\nu_t$ $(t\ge 2)$ from
$$
\phi\sim \mathrm{IG}\left(1,1+\frac{1}{\tau^2}\right),\quad\nu_t\sim \mathrm{IG}\left(1,1+\frac{1}{u_t^2}\right)
$$
\item[--] Sample $\rho_t$ ($t=1,\dots,T$) from the grid-based distribution proportional to
$$
|(1-\rho_t)I+\rho_t(D-W)|^{1/2}\exp\left(-\frac{1}{2\tau^2u_t^2}\eta_t^\top\{(1-\rho_t)I+\rho_t(D-W)\}\eta_t\right)\pi(\rho_t)
$$
at the 99 grid points $\{0.01,0.02,\dots,0.99\}$, where $\tau^2=1$ for $t=1$. 
\item[--] Sample $\sigma^2$ from 
\begin{equation*}
    \sigma^2\mid y,\theta, \lambda, \tau, \rho\sim \mathrm{IG}\left(\frac{nT}{2}+a_{\sigma^2},\sum_{t=1}^T\sum_{i=1}^n\frac{(y_{it}-\theta_{it})^2}{2}+b_{\sigma^2}\right).
\end{equation*}
\end{itemize}
The default values of the hyperparameters are $a_{u_1}=b_{u_1}=0.1$ and $a_{\sigma^2}=b_{\sigma^2}=0.1$. Sampling $\rho_t$ for $t=1,\dots,T$ is computationally inexpensive because each $\rho_t$ is a scalar parameter restricted to the interval $[0,1)$. In practice, we sample $\rho_t$ from a prespecified discrete grid, and a similar discrete sampling strategy for the spatial dependence parameter has also been considered by \cite{gelfand2003proper}.

\section{Simulation study}

To assess the performance of the proposed method, we give a simulation study in this section. This paper focuses on spatiotemporal denoising and the recovery of interpretable spatiotemporal trends, and its motivation is similar to that of nonparametric smoothing methods such as splines. Therefore, in the simulation study, we assume smooth and interpretable underlying spatiotemporal trends and evaluate the ability of each method to recover these trends.

\subsection{Simulation settings}

\begin{figure}[ht]
    \centering
    \includegraphics[width=14cm]{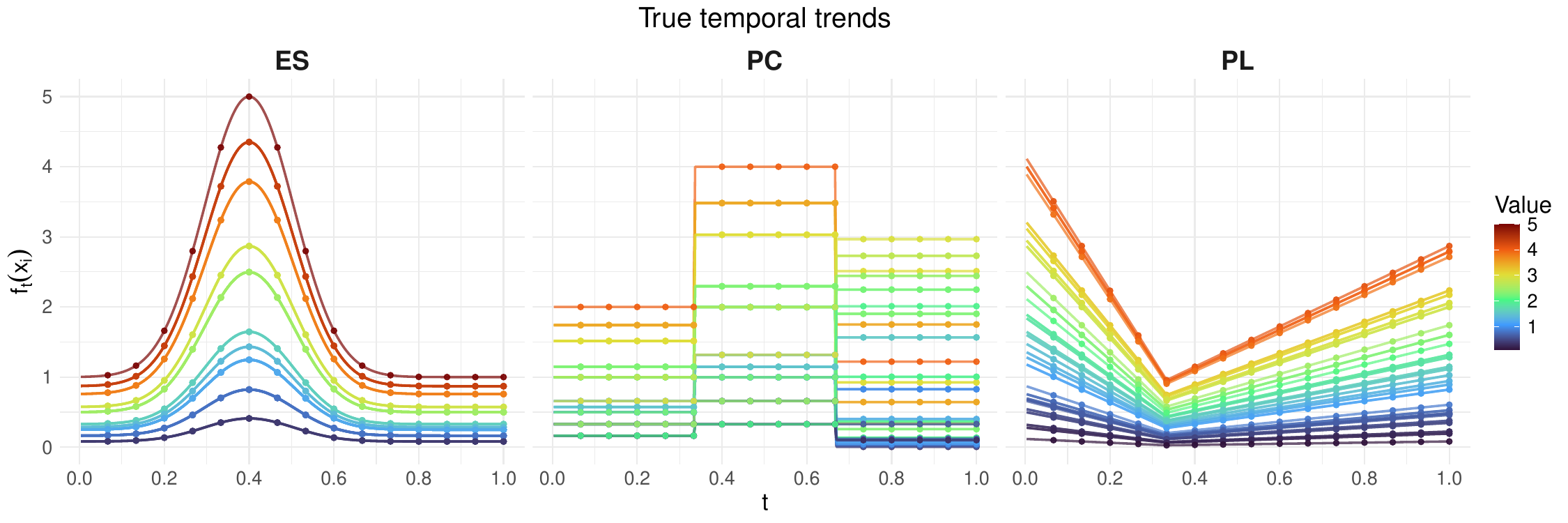}
    \caption{The true underlying functions $f_t(x_i)$ of ES, PC and PL from left to right under $n=49$ and $T=15$ are colored by ordering of the maximum value of each trend. Each trend represents the temporal trends of each area. }
    \label{fig:sim-true-temporal-trends}
\end{figure}

We consider a two-dimensional lattice graph with $n=d\times d$ vertices (areas). Two areas are defined to be adjacent if their corresponding lattice points are neighbors, and the adjacency matrix $W\in\mathbb{R}^{n\times n}$ is constructed from this two-dimensional lattice graph. The data are generated according to
\begin{equation*}
y_{it} = f_t(x_i) + \epsilon_{it},\quad i=1,\dots,n,\ t\in\{m/T\in [0,1]\mid m=1,\dots,T\}
\end{equation*}
where $T\in\{15,30\}$ and $\epsilon_{it}\overset{\mathrm{i.i.d.}}{\sim}N(0,0.1)$. The vector $x_i=(x_{i1},x_{i2})^\top\in D$ denotes the spatial coordinate at which $y_{it}$ is observed, where
\begin{equation*}
    D=
    \left\{
        \left(\frac{j}{d-1},\frac{k}{d-1}\right):
        j,k=0,1,\dots,d-1
    \right\}
    \subset [0,1]^2
\end{equation*}
is the set of regularly spaced grid points on the unit square. We consider two lattice sizes, $n=49$ ($d=7$) and $n=121$ ($d=11$). The function $f_t(x_i)$ denotes the underlying spatiotemporal trend evaluated at location $x_i$ and time $t$, and is defined as follows:
$$
f_t(x_i)=h_t\exp\left\{-20(x_i-\mu_t)^\top(x_i-\mu_t)\right\}.
$$
We consider two settings for $\mu_t$ and $h_t$ as follows: 
\begin{itemize}
\item[--] Exponential smooth (ES) 
\begin{equation*}
    \mu_t=(0.5,0.5)^\top,\quad h_t=1+4\exp\{-45(t-0.5)^2\}.
\end{equation*}
\item[--] Piecewise constant (PC)
\begin{equation*}
\mu_t=
    \begin{cases}
    (0.5,0.5)^\top & 0\le t \le 2/3\\
    (0.8,0.8)^\top & 2/3< t\le 1
    \end{cases}, \quad h_t=
    \begin{cases}
        2 & 0\le t \le 1/3\\
        4 & 1/3< t \le 2/3\\
        3 & 2/3< t\le 1
    \end{cases}.
\end{equation*}
\item[--] Piecewise linear (PL)
\begin{equation*}
    \mu_t=(0.4,0.4)^\top,\quad h_t=1+(-10t+10/3)\cdot1_{\{0<t\le1/3\}}+(3t-1)\cdot1_{\{0<t\le1/3\}}.
\end{equation*}
\end{itemize}
Such an exponential spatial setting $f_t(x_i)$ is also used in \cite{onizuka2024locally}, and the piecewise polynomial temporal settings such as PC and PL are also considered in several studies as examples \citep[e.g.][]{tibshirani2014adaptive, faulkner2018locally}. The true spatiotemporal trends are shown in Figure~\ref{fig:sim-true-temporal-trends}, which displays the temporal trajectories for all $n=49$ areas under each scenario. Although there are 49 lines in total, some are identical by construction, while others visually overlap. The ES function does not exhibit any abrupt temporal changes, and the PC function contains three piecewise-constant segments due to the definitions of $\mu_t$ and $h_t$. To evaluate the performance of the proposed locally adaptive method, we compare it with the following existing methods:
\begin{itemize}
\item GLS: the proposed global local shrinkage model. 
\item GS: the global shrinkage model, which corresponds to setting temporal correlation $u_t=1$ ($t\ge 2$) in the proposed model. 
\item TGL: non-spatial (temporal) global local shrinkage model, which corresponds to setting spatial correlation $\rho_t=0$ in the proposed model. 
\item AR: autoregressive model proposed by \cite{rushworth2014spatio}. This method is implemented by the \texttt{ST.CARar} function of the \texttt{CARBayesST} package in R software \citep[see e.g.][]{lee2018spatio}.
\item IND: independent CAR model for each time $t$. This method is implemented by the \texttt{S.CARleroux} function of the \texttt{CARBayes} package in R software.
\end{itemize}
We apply the first-order models in the PC scenario and second-order models in ES and PL scenarios for GLS, GL, and AR models. We generated 5000 posterior samples and discarded the first 2000 samples as burn-in for all methods. 
To evaluate two methods, we calculate the mean squared error (MSE), mean absolute deviation (MAD), average length of 95\% credible intervals (AL) and coverage probability of 95\% credible intervals (CP), which are defined as follows:
\begin{align*}
    &\mathrm{MSE}=\frac{1}{nT}\sum_{t=1}^T\sum_{i=1}^n \left(f_t(x_i)-\hat{\theta}_{it}\right)^2,\quad \mathrm{MAD}=\frac{1}{nT}\sum_{t=1}^T\sum_{i=1}^n \left|f_t(x_i)-\hat{\theta}_{it}\right|,\\
    &\mathrm{AL}=\frac{1}{nT}\sum_{t=1}^T\sum_{i=1}^n \hat{\theta}^{(0.975)}_{it}-\hat{\theta}^{(0.025)}_{it},\quad
    \mathrm{CP}=\frac{1}{nT}\sum_{t=1}^T\sum_{i=1}^n I(\hat{\theta}^{(0.025)}_{it}\le f_t(x_i)\le \hat{\theta}^{(0.975)}_{it}),
\end{align*}
where $\hat{\theta}_{it}$ and $\hat{\theta}^{(q)}_{it}$ represent a point estimate of the $\theta_{it}$ and $q$-quantiles of the posterior distribution of $\theta_{it}$, respectively. 
We generate 100 datasets for each scenario and compute the average of these criteria over 100 repetitions. Note that the other simulation settings are also provided in Appendix~\ref{Appendix-sim}.

\subsection{Simulation results}

\begin{figure}
    \centering
    \includegraphics[width=14cm]{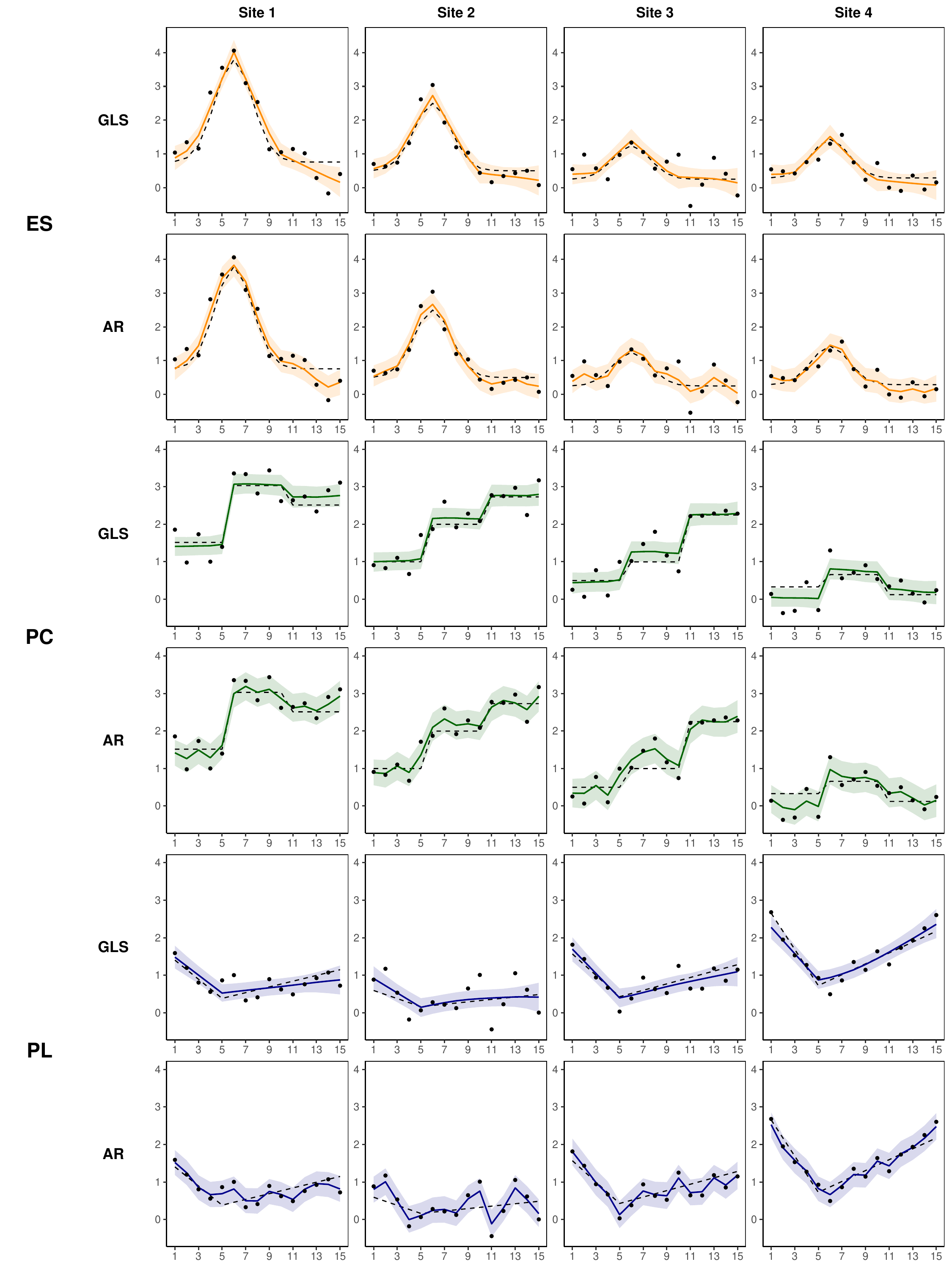}
    \caption{The observed data (black points), true trends (dashed lines), and estimated temporal trends (solid lines) obtained by the proposed GLS method (odd rows) and the AR method (even rows) for four selected areas under the ES (orange), PC (green), and PL (blue) scenarios with $n=49$ and $T=15$.}
    \label{fig:sim-est-temporal-trends}
\end{figure}

\begin{figure}
    \centering
    \includegraphics[width=14cm]{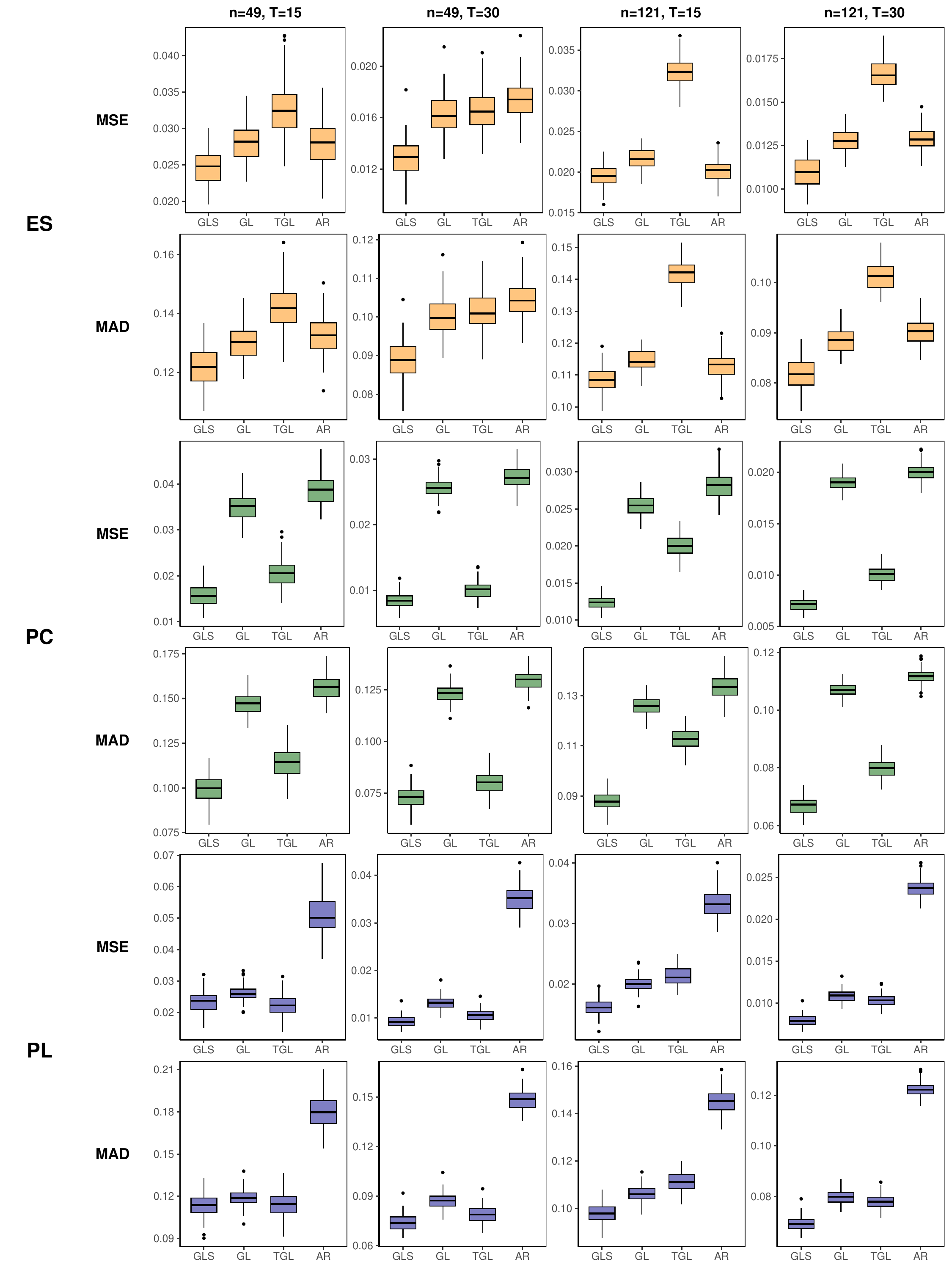}
    \caption{Boxplots of the MSE and MAD values for the ES (orange), PC (green), and PL (blue) scenarios.}
    \label{fig:sim-MSEMAD}
\end{figure}

The simulated data and the corresponding true and estimated temporal trends for selected 4 areas are shown in Figure~\ref{fig:sim-est-temporal-trends}. For simplicity, we only visualize the results of the proposed GLS method and the existing AR method. 
Although the temporal trends vary across areas, the proposed GLS method successfully captures the underlying patterns. In particular, it accurately recovers the piecewise-constant trends in the PC scenario and provides interpretable estimates for all considered true functions.
The boxplots of MSE and MAD values over 100 replications are summarized in Figure~\ref{fig:sim-MSEMAD}. The proposed method achieves the smallest MSE and MAD in almost all scenarios, indicating its superior performance in terms of point estimation. Moreover, the MSE and MAD generally decrease as the number of areas $n$ or the number of time points $T$ increases. The TGL method also performs well for large $T$ or in a PC scenario, but its performance does not improve substantially as $n$ increases because it cannot exploit the spatial neighborhood information. Therefore, the results suggest that the proposed GLS method successfully borrows information from neighboring areas and adjacent time points, leading to a stable and accurate estimation. Note that the results of the IND method are not shown in Figure~\ref{fig:sim-MSEMAD} for clarity and the averaged MSE and MAD including the IND method are also summarized in Appendix~\ref{Appendix-sim}. 
The coverage probabilities (CPs) and average lengths (ALs) of the 95\% credible intervals are summarized in Table~\ref{Table:sim-CPAL}. The CPs of the proposed, GS, and AR methods are generally close to the nominal level of 0.95. In addition, the proposed method tends to yield shorter credible intervals, as indicated by its smaller AL values, while maintaining appropriate coverage probabilities. Because the IND method does not borrow information from spatial and temporal neighborhoods, it has the worst performance regardless of scenario. 
Overall, the proposed approach tends to perform well not only for piecewise-polynomial temporal trends but also for smooth temporal trends, such as those in the ES scenario.

\begin{table}[htbp]
\caption{The averaged CP and AL over 100 Monte Carlo simulations.}
\begin{center}
\begin{tabular}{ccc|ccccc}
  \toprule
\multicolumn{8}{c}{ES}\\
  \midrule
$n$ & $T$ &  & GLS & GS & TGL & AR & IND \\ 
  \midrule
     \multirow{4}{*}{49} & \multirow{2}{*}{15} & CP & 0.936 & 0.941 & 0.942 & 0.961 & 0.575 \\ 
   & & AL & 0.583 & 0.638 & 0.687 & 0.692 & 0.612 \\ 
   & \multirow{2}{*}{30} & CP & 0.948 & 0.947 & 0.951 & 0.952 & 0.579 \\ 
   & & AL & 0.442 & 0.495 & 0.506 & 0.551 & 0.617 \\ 
     \multirow{4}{*}{121} & \multirow{2}{*}{15} & CP & 0.944 & 0.950 & 0.940 & 0.977 & 0.578 \\ 
   & & AL & 0.535 & 0.576 & 0.683 & 0.638 & 0.615 \\ 
    & \multirow{2}{*}{30} & CP & 0.948 & 0.952 & 0.943 & 0.973 & 0.579 \\ 
   & & AL & 0.404 & 0.448 & 0.491 & 0.521 & 0.621 \\ 
\midrule
\midrule
\multicolumn{8}{c}{PC}\\
  \midrule
$n$ & $T$ &  & GLS & GS & TGL & AR & IND \\ 
  \midrule
     \multirow{4}{*}{49} & \multirow{2}{*}{15} & CP & 0.968 & 0.952 & 0.960 & 0.942 & 0.291 \\ 
   & & AL  & 0.537 & 0.747 & 0.598 & 0.746 & 0.450 \\ 
   & \multirow{2}{*}{30} & CP & 0.971 & 0.959 & 0.968 & 0.961 & 0.291 \\ 
   & & AL & 0.406 & 0.667 & 0.436 & 0.692 & 0.452 \\ 
     \multirow{4}{*}{121} & \multirow{2}{*}{15} & CP & 0.972 & 0.965 & 0.960 & 0.962 & 0.279 \\ 
   & & AL& 0.484 & 0.678 & 0.585 & 0.695 & 0.431 \\ 
    & \multirow{2}{*}{30} & CP & 0.969 & 0.966 & 0.959 & 0.971 & 0.282 \\
   & & AL & 0.362 & 0.601 & 0.414 & 0.627 & 0.435 \\ 
\midrule
\midrule
\multicolumn{8}{c}{PL}\\
  \midrule
$n$ & $T$ &  & GLS & GS & TGL & AR & IND \\ 
  \midrule
     \multirow{4}{*}{49} & \multirow{2}{*}{15} & CP & 0.919 & 0.914 & 0.955 & 0.874 & 0.344 \\ 
   & & AL & 0.499 & 0.525 & 0.566 & 0.686 & 0.462 \\ 
   & \multirow{2}{*}{30} & CP & 0.955 & 0.939 & 0.959 & 0.944 & 0.340 \\ 
   & & AL & 0.371 & 0.421 & 0.404 & 0.717 & 0.463 \\ 
     \multirow{4}{*}{121} & \multirow{2}{*}{15} & CP & 0.940 & 0.930 & 0.952 & 0.942 & 0.291 \\ 
   & & AL & 0.464 & 0.496 & 0.552 & 0.690 & 0.391 \\ 
    & \multirow{2}{*}{30} & CP & 0.954 & 0.945 & 0.955 & 0.969 & 0.290 \\ 
   & & AL & 0.344 & 0.392 & 0.392 & 0.666 & 0.396 \\ 
  \bottomrule
\end{tabular}
\label{Table:sim-CPAL}
\end{center}
\end{table}

\section{Applications}

\subsection{Housing market data}
\begin{figure}
    \centering
    \includegraphics[width=14cm]{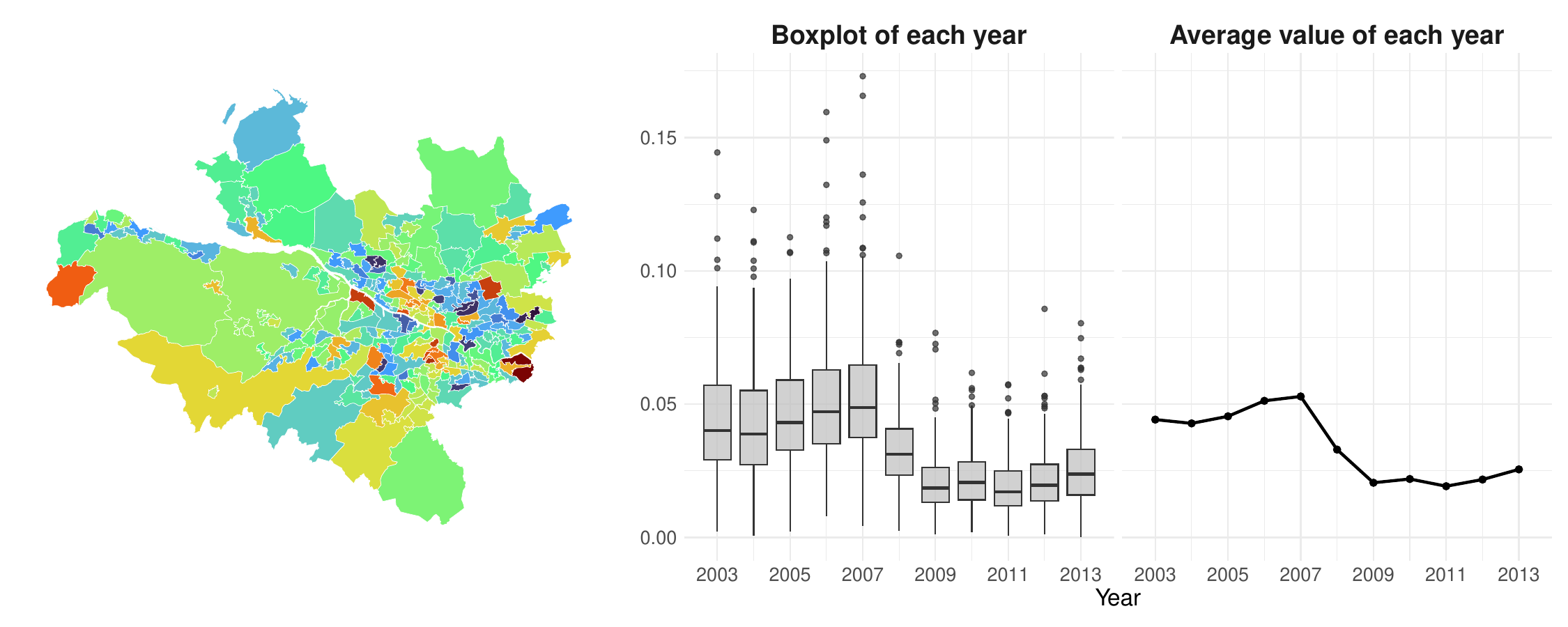}
    \caption{The averaged spatial trend over 11 years (left), boxplots of the raw rates from 2003 to 2013 (center), and the mean temporal trend from 2003 to 2013 (right).}
    \label{fig:housing-dataplot}
\end{figure}

We adopt the property sales data available in the {\tt salesdata} object of the {\tt CARBayesdata} R package. \cite{lee2018spatio} also analyzed the state of the housing market, specifically property sales, measured as the proportion of properties sold relative to the total number of properties. This rate is used as an approximate measure of property market activity. The data consist of annual observations for the 271 Intermediate Zones that comprise the Greater Glasgow and Clyde Health Board in Scotland, covering the period from 2003 to 2013, namely $n=271$ and $T=11$ in our setting. The averaged spatial trend over 11 years is shown in the left panel of Figure~\ref{fig:housing-dataplot}. The areas shaded in red, representing higher rates, are scattered across the study region, whereas the blue and green areas, representing relatively lower rates, appear to display some degree of spatial dependence. The Moran's I statistic of the averaged spatial trend is about 0.30. As illustrated by the boxplots and the average temporal trend in the center and right panels of Figure~\ref{fig:housing-dataplot}, respectively, property sales generally increased until 2007 and then declined sharply between 2007 and 2008. \cite{lee2018spatio} also identified this abrupt change as an important feature of the data. We therefore aim to quantify such temporal changes at the area level during a period that includes the global financial crisis that began in late 2007. To capture such temporal change, we apply the first-order proposed model and the AR (1) model. We generated 20000 posterior samples, and only every 5th scan was saved (thinning) after discarding the first 2000 samples as burn-in for all methods.

\begin{figure}
    \centering
    \includegraphics[width=14cm]{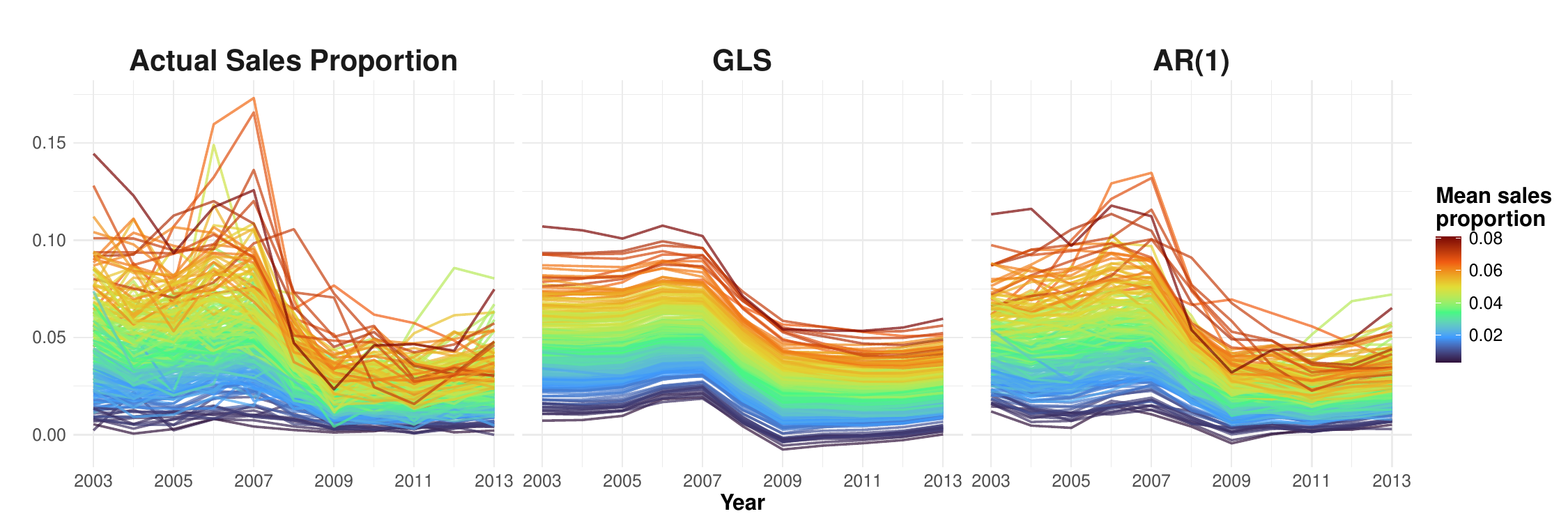}
    \caption{The actual data (left), the estimated temporal trends by the proposed GLS (center) and AR (right) from 2003 to 2013 for 271 Intermediate Zones.}
    \label{fig:housing-actual-est-trend}
\end{figure}

Figure~\ref{fig:housing-actual-est-trend} shows, from left to right, the observed data and the temporal trends estimated by the proposed GLS and AR(1) methods. The trajectories are colored according to the average value for each of the 271 Intermediate Zones over the observation period. The proposed GLS method successfully captures abrupt changes around the 2008 global financial crisis, while also producing piecewise-constant temporal trends for individual areas.
We also report the posterior means of the local parameters $u_t$ and $\rho_t$ for the proposed method in Table~\ref{tab:housing-rho-u}. The local scale parameter $u_t$ ($t\ge 2$) controls the degree of shrinkage of the temporal difference, and hence a large value of $u_t$ indicates a potentially substantial temporal change at time $t$ ($t\ge 2$) as mentioned in Remark~\ref{rem-shrinkage-factor}. In contrast, $\rho_t$ ($t\ge 2$) controls the strength of spatial dependence in the temporal differences. When the spatial changes occur simultaneously over the entire study region, the estimated value of $\rho_t$ tends to become large. Thus, $\rho_t$ can be interpreted as capturing the degree of spatial coherence in the temporal evolution and is estimated to be smaller if the spatial trend does not change between $t-1$ and $t$. 
The estimated values of $u_t$ and $\rho_t$ suggest that substantial changes in the temporal trends occurred around 2008 and 2009, which is consistent with the abrupt decline in property sales during the global financial crisis.

\begin{table}[htbp]
\caption{Posterior means of $\rho_t$ and $u_t$. Values satisfying $\hat{\rho}_t>0.9$ or $\hat{u}_t>1$ are displayed in bold.}
\begin{center}
\begin{tabular}{rrrrrrrrrrrr}
  \toprule
& 2003 & 2004 & 2005 & 2006 & 2007 & 2008 & 2009 & 2010 & 2011 & 2012 & 2013 \\ 
  \midrule
$\hat{\rho}_t$& {\bf 0.98} & 0.52 & 0.53 & {\bf 0.99} & 0.55 & {\bf 0.98} & {\bf 0.99} & 0.51 & 0.52 & 0.52 & {\bf 0.96} \\ 
  $\hat{u}_t$ & 0.04 & 0.57 & 0.43 & {\bf 2.54} & 0.38 & {\bf 8.77} & {\bf 3.49} & 0.43 & 0.42 & 0.41 & {\bf 1.39} \\ 
   \bottomrule
\end{tabular}
    \label{tab:housing-rho-u}
\end{center}
\end{table}

\subsection{Italian unemployment data }

\begin{figure}
    \centering
    \includegraphics[width=14cm]{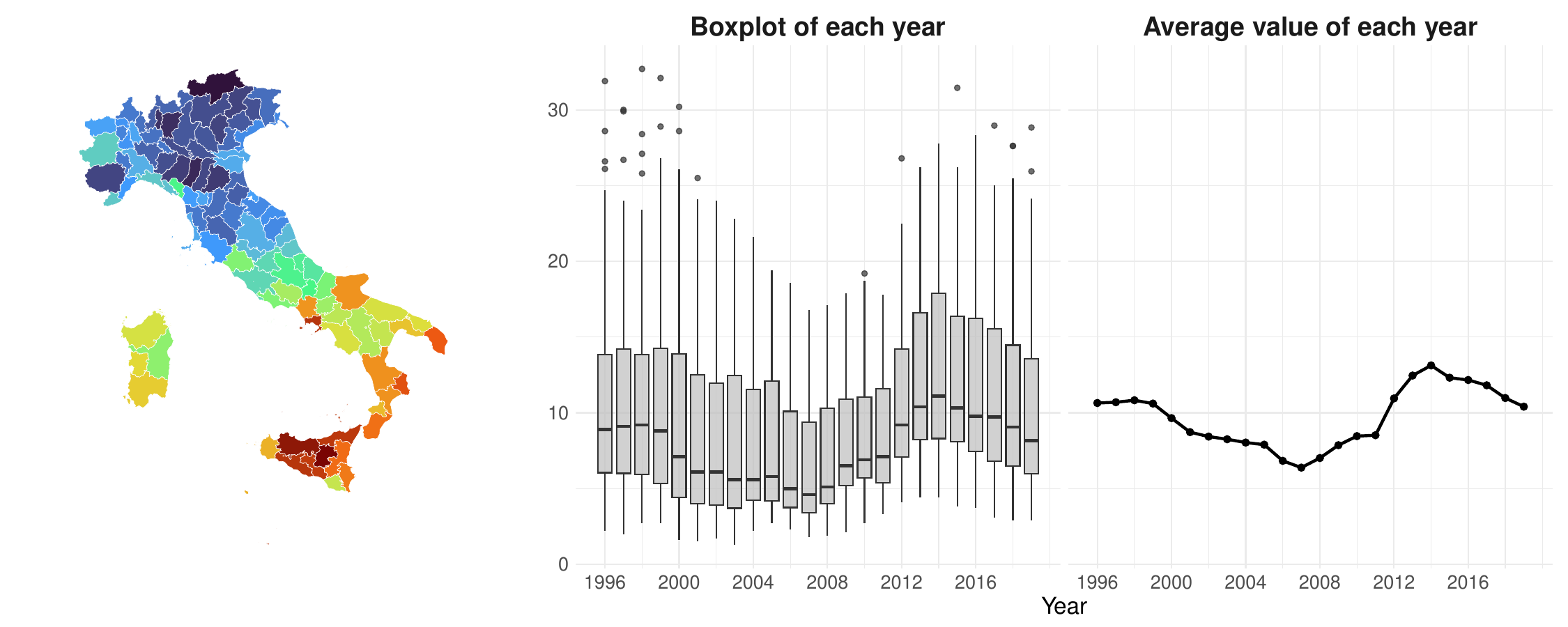}
    \caption{The averaged spatial trend over 24 years (left), boxplots of the raw unemployment rates from 1996 to 2019 (center), and the mean temporal trend from 1996 to 2019 (right).}
    \label{fig:unemp-dataplot}
\end{figure}

\begin{figure}
    \centering
    \includegraphics[width=14cm]{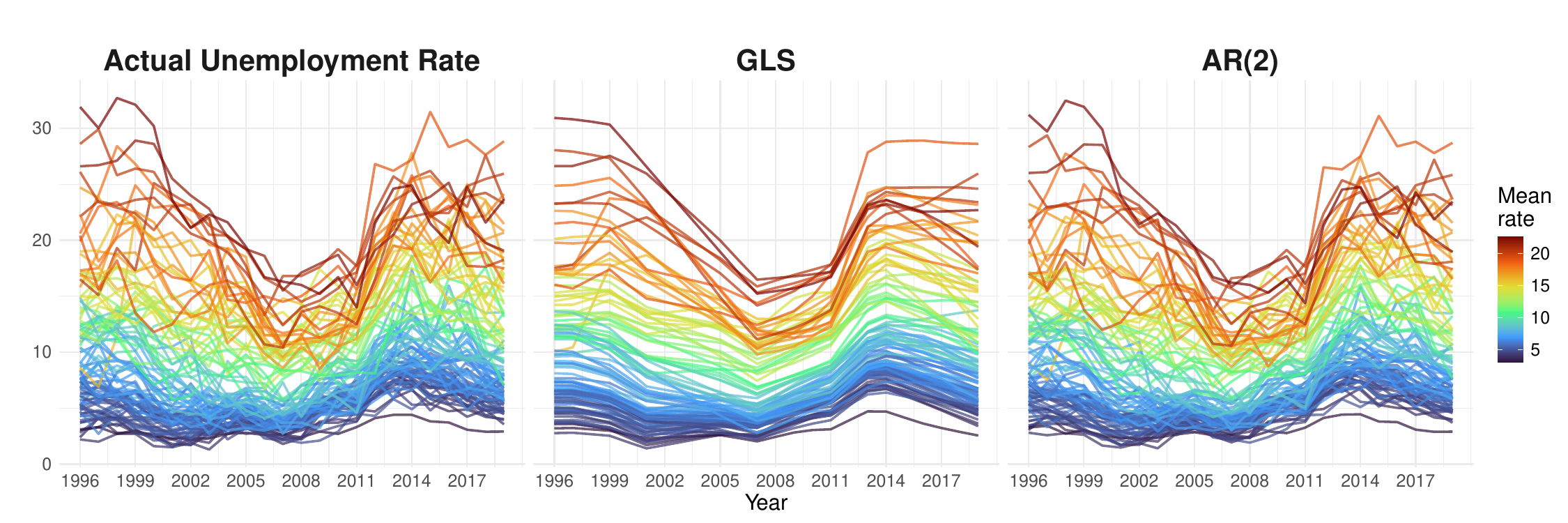}
    \caption{The observed data (left) and the temporal trends estimated by the proposed GLS method (center) and the AR method (right) from 1996 to 2019 for 103 areas.}
    \label{fig:unemp-actual-est-trend}
\end{figure}

The proposed method is also applied to the Italian unemployment data available as the {\tt unemp\_it} object in the {\tt pspatreg} R package, which is considered by \cite{minguez2022introduction}. We focus on the unemployment rate $y_{it}$ for 103 Italian NUTS-3 provinces observed annually from 1996 to 2019, where $i=1,\dots,103$ and $t=1,\dots,24$. The main response variable is {\tt unrate}, the unemployment rate in percentage terms for each province and year. The spatial trend averaged over the 24-year period, the yearly boxplots, and the annual average unemployment rates are shown in Figure~\ref{fig:unemp-dataplot}. These plots indicate substantial spatial dependence as well as temporal variation in the unemployment rates. The unemployment rates exhibit a pronounced spatial pattern reflecting the well-known north-south divide in Italy. Provinces in northern and some central regions colored blue tend to have relatively low unemployment rates, whereas high unemployment rates colored red are concentrated in southern provinces and the islands. This visual impression is supported by a Moran's I statistic of the averaged spatial trend about 0.89. \cite{mozdzen2022spatial} also analyzed the same data using explanatory variables. As shown in the left panel of Figure~\ref{fig:unemp-actual-est-trend}, most areas exhibit broadly similar temporal patterns, characterized by a U-shaped trend with relatively low unemployment rates around 2006--2011. In addition, the annual average unemployment rate shown in the right panel of Figure~\ref{fig:unemp-dataplot} generally decreases from 1999 to 2007 and again after 2014. For posterior inference, we generated 20,000 MCMC samples for each method and discarded the first 2,000 samples as burn-in, and then we retained every fifth sample to reduce the storage and dependence of the MCMC output. We compare the proposed second-order model with the AR(2) model.

The temporal trends estimated by the GLS and AR(2) methods for all areas are shown in the center and right panels of Figure~\ref{fig:unemp-actual-est-trend}, respectively. As shown in Figure~\ref{fig:unemp-actual-est-trend}, the GLS method yields smooth and interpretable temporal trends that clearly reflect the overall decreases in unemployment rates until 2007 and again after 2014. Moreover, most areas have similar temporal trends, and the spatial pattern of the unemployment rate does not differ substantially across years, as the relative ordering of the colors remains largely unchanged. This implies that the potential gap in the unemployment rate between northern and southern Italy persists over time. In contrast, the AR(2) method produces more irregular temporal trends that appear to be more strongly influenced by noise in the observed data. A similar tendency was also observed in the simulation study.

\section{Concluding remarks}\label{sec:concluding}

The main contribution of this study is to model temporal changes in the latent spatial surface directly, rather than imposing a globally smooth autoregressive evolution on the latent effects themselves. By placing spatially structured global-local shrinkage priors on temporal differences, the proposed model can adaptively distinguish periods of stability from periods of structural change. This feature is particularly useful for areal spatiotemporal data in which regional outcomes evolve smoothly during some periods but exhibit abrupt changes due to external shocks or interventions. Through simulation studies and real data analysis, we demonstrate that the proposed method can improve the estimation of spatiotemporal trends and provide more flexible inference than standard AR-type spatiotemporal CAR models.

There are several topics for future work. First, the proposed spatiotemporal CAR prior is also applied to small area estimation by introducing a covariate term and modifying the likelihood. 
Although the paper deals with a Gaussian likelihood for simplicity, other likelihoods such as Poisson, negative binomial, and so on are also considered by adopting knowledge of the generalized linear models. The proposed model assumes that the underlying temporal trends of all areas share the jump/change points. To deal with the heterogeneity of temporal trends between areas, introducing a local parameter for each area and time is also considered \citep[e.g.][]{mozdzen2022spatial,tang2023global,nishina2026global}. In this regard, not only temporal local adaptivity but also spatial local adaptivity is an important problem \citep[see also][]{onizuka2024locally}. Furthermore, time-varying parameter models with time-varying regression coefficients are also of interest, and can be developed by adopting the multivariate formulation and shrinkage priors.

\section*{Acknowledgments}

The authors' research was supported in part by JSPS KAKENHI Grant Numbers 25K23108, 25K07131 and 26K00322 from the Japan Society for the Promotion of Science.

\bibliographystyle{chicago}
\bibliography{refs}

\newpage

\appendix

\section{Additional numerical example}
\label{Appendix-sim}

\begin{figure}[ht]
    \centering
    \includegraphics[width=10cm]{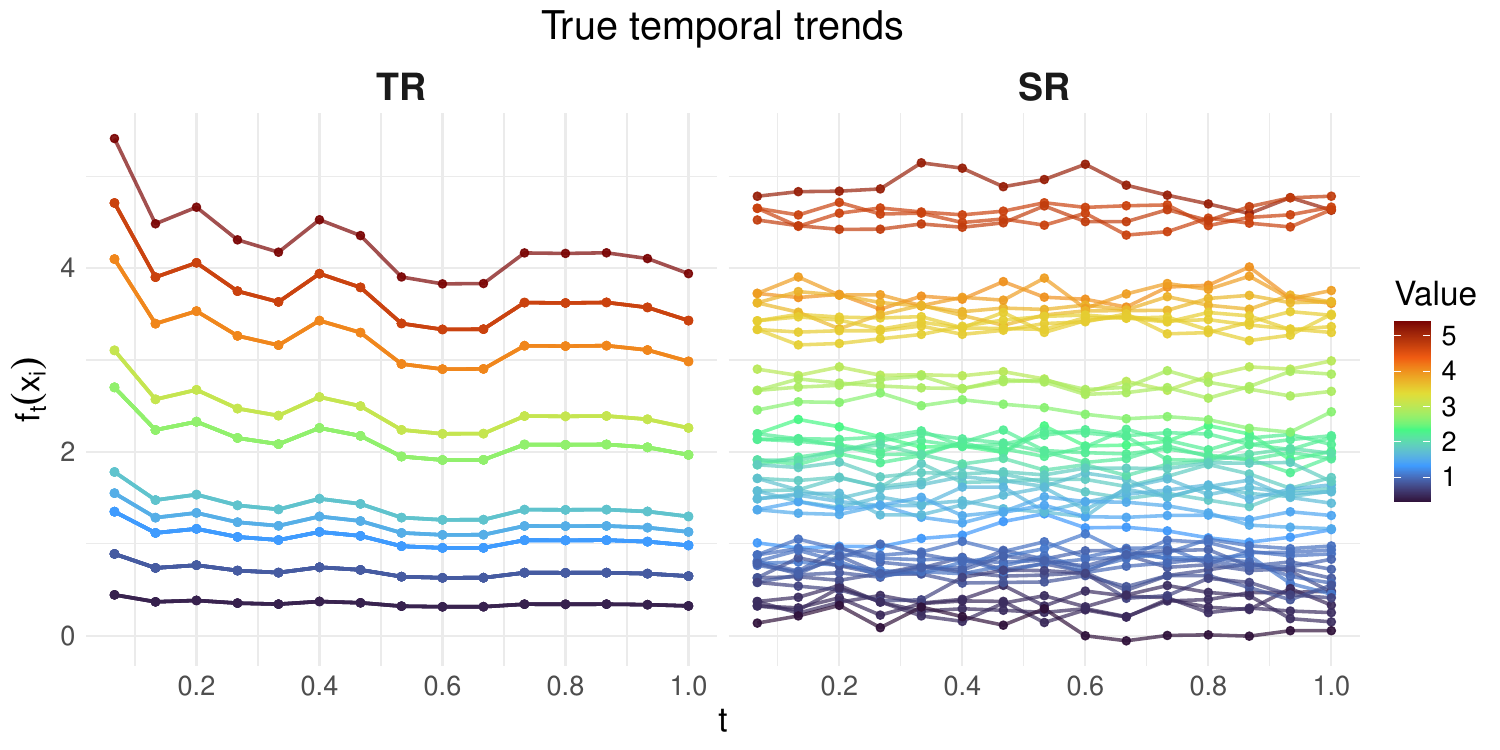}
    \caption{The true underlying functions $f_t(x_i)$ under the TR and SR scenarios, shown from left to right, respectively, with $n=49$ and $T=15$. The curves are colored according to the ordering of the maximum value of each temporal trend. Each curve represents the temporal trend for one area.}
    \label{fig:sim-true-temporal-trends-appendix}
\end{figure}

Studies based on smoothing methods such as splines typically consider settings in which the underlying trend is interpretable or smooth, as in the main manuscript. In this section, however, we additionally consider a setting in which the underlying spatiotemporal trends are rough. The data generating process except the underlying functions $f_t$ is same as the setting of the main manuscript, and $f_t(\cdot)$ is defined as follows: 
\begin{itemize}
\item[--] Temporal roughness (TR)
\begin{align*}
&f_t(x_i)=h_t\exp\left\{-20(x_i-\mu_t)^\top(x_i-\mu_t)\right\},\quad \mu_t=(0.5,0.5)^\top,\\
&h_t=3 + \exp(g_t),\quad g_t=0.5g_{t-1}+\xi_t,\quad \xi_t\sim N(0, 0.2).
\end{align*}
\item[--] Spatiotemporal roughness (SR)
\begin{align*}
&f_1(x_i)=5\exp\left\{-20(x_i-\mu_1)^\top(x_i-\mu_1)\right\},\quad \mu_1=(0.4,0.4)^\top,\\
&f_t(x_i)=f_1(x_i)+0.5(f_{t-1}(x_i)-f_1(x_i))+\xi_t,\quad \xi_t\sim N(0, 0.1)\quad t\ge 2.
\end{align*}
\end{itemize}
Although the settings are similar to the scenarios in the main manuscript, they are based on an autoregressive process. 
The true spatiotemporal trends are shown in Figure~\ref{fig:sim-true-temporal-trends-appendix}, which displays the temporal trajectories for all $n = 49$ areas under each scenario. As shown in Figure~\ref{fig:sim-true-temporal-trends-appendix}, the functions are not smooth and rough unless the main manuscript. We compare the proposed method with the same methods as those considered in the main manuscript and apply a first-order model to all spatiotemporal models. Performance is evaluated in terms of MSE, MAD, CP, and AL, and the other settings remain the same.

We summarize the averaged MSE and MAD over 100 Monte Carlo simulations including ES, PC and PL scenarios of the main manuscript in Tables~\ref{Table:sim-MSE} and \ref{Table:sim-MAD}. Although the proposed GLS method yields the smallest values of MSE and MAD in most scenario, there are no substantial differences among the GLS, GS, and TGL methods. In addition, the GL method provides suitable estimates in the TR and SR scenarios compared with those in the other scenarios. The CP and AL values are reported in Table~\ref{Table:sim-appendix-CPAL}. In terms of uncertainty quantification, the CP values of the AR method are close to the nominal coverage level of 0.95 for all scenarios. However, unlike the results for point estimation, the GLS, GS, and TGL methods underperform in the SR scenario in terms of uncertainty quantification, with shorter AL values accompanied by lower CP values than those of the AR method. It suggests that, under such misspecification, we have to construct general posterior based on a quadratic loss and calibrate the posterior distributions.

\begin{table}[htbp]
\caption{The averaged MSE over 100 Monte Carlo simulations. The minimum value is displayed in bold.}
\begin{center}
\begin{tabular}{ccc|ccccc}
  \toprule
\multicolumn{8}{c}{MSE}\\
  \midrule
$f_t(\cdot)$ & $n$ & $T$ & GLS & GS & TGL & AR & IND \\ 
  \midrule
     \multirow{4}{*}{ES} & \multirow{2}{*}{49} & 15 & {\bf 0.025} & 0.028 & 0.033 & 0.028 & 0.778 \\ 
   & & 30  & {\bf 0.013} & 0.016 & 0.017 & 0.017 & 0.767 \\
   & \multirow{2}{*}{121} & 15  & {\bf 0.020} & 0.022 & 0.032 & 0.020 & 0.815 \\ 
   & & 30 & {\bf 0.011} & 0.013 & 0.017 & 0.013 & 0.817 \\ 
   \multirow{4}{*}{PC} & \multirow{2}{*}{49} & 15 & {\bf 0.016} & 0.035 & 0.021 & 0.039 & 1.053 \\ 
   & & 30 & {\bf 0.009} & 0.026 & 0.010 & 0.027 & 1.042 \\ 
   & \multirow{2}{*}{121} & 15& {\bf 0.012} & 0.025 & 0.020 & 0.028 & 1.097 \\ 
   & & 30  & {\bf 0.007} & 0.019 & 0.010 & 0.020 & 1.103 \\ 
   \multirow{4}{*}{PL} & \multirow{2}{*}{49} & 15 & 0.023 & 0.026 & {\bf 0.022} & 0.051 & 0.354 \\ 
   & & 30 & {\bf 0.009} & 0.013 & 0.010 & 0.035 & 0.362 \\ 
   & \multirow{2}{*}{121} & 15  & {\bf 0.016} & 0.020 & 0.021 & 0.033 & 0.384 \\ 
   & & 30 & {\bf 0.008} & 0.011 & 0.010 & 0.024 & 0.385 \\ 
     \multirow{4}{*}{TR} & \multirow{2}{*}{49} & 15 & 0.013 & {\bf 0.013} & 0.015 & 0.033 & 0.114 \\ 
   & & 30  & {\bf 0.010} & 0.010 & 0.012 & 0.024 & 0.116 \\ 
   & \multirow{2}{*}{121} & 15 & {\bf 0.010} & 0.010 & 0.016 & 0.023 & 0.105 \\ 
   & & 30 & {\bf 0.008} & 0.008 & 0.013 & 0.018 & 0.110 \\ 
      \multirow{4}{*}{SR} & \multirow{2}{*}{49} & 15 & {\bf 0.015} & 0.015 & 0.015 & 0.040 & 0.152 \\ 
   & & 30  & {\bf 0.014} & 0.014 & 0.014 & 0.031 & 0.142 \\  
   & \multirow{2}{*}{121} & 15 &{\bf 0.015} & 0.015 & 0.016 & 0.030 & 0.114 \\ 
   & & 30 & 0.014 & {\bf 0.014} & 0.014 & 0.025 & 0.119 \\ 
   \bottomrule
\end{tabular}
\label{Table:sim-MSE}
\end{center}
\end{table}

\begin{table}[htbp]
\caption{The averaged MAD over 100 Monte Carlo simulations. The minimum value is displayed in bold.}
\begin{center}
\begin{tabular}{ccc|ccccc}
  \toprule
   \multicolumn{8}{c}{MAD}\\
  \midrule
$f_t(\cdot)$ & $n$ & $T$ & GLS & GS & TGL & AR & IND \\ 
  \midrule
     \multirow{4}{*}{ES} & \multirow{2}{*}{49} & 15 & {\bf 0.122} &  0.130 & 0.142 & 0.133 & 0.528 \\ 
   & & 30 & {\bf 0.089} & 0.100 & 0.101 & 0.104 & 0.525 \\ 
   & \multirow{2}{*}{121} & 15 & {\bf 0.108} & 0.115 & 0.142 & 0.113 & 0.551 \\ 
   & & 30 & {\bf 0.082} & 0.089 & 0.101 & 0.090 & 0.552 \\ 
\multirow{4}{*}{PC} & \multirow{2}{*}{49} & 15 & {\bf 0.099} & 0.147 & 0.114 & 0.156 & 0.738 \\ 
   & & 30  & {\bf 0.073} & 0.123 & 0.080 & 0.130 & 0.735 \\ 
   & \multirow{2}{*}{121} & 15 & {\bf 0.088} & 0.126 & 0.113 & 0.133 & 0.755 \\
   & & 30& {\bf 0.067} & 0.107 & 0.080 & 0.112 & 0.757 \\ 
   \multirow{4}{*}{PL} & \multirow{2}{*}{49} & 15 & {\bf 0.114} & 0.119 & 0.115 & 0.180 & 0.454 \\ 
   & & 30  & {\bf 0.074} & 0.087 & 0.079 & 0.148 & 0.461 \\ 
   & \multirow{2}{*}{121} & 15 & {\bf 0.098} & 0.106 & 0.112 & 0.145 & 0.472 \\ 
   & & 30 & {\bf 0.069} & 0.080 & 0.078 & 0.122 & 0.476 \\ 
    \multirow{4}{*}{TR} & \multirow{2}{*}{49} & 15 & 0.088 & {\bf 0.088} & 0.095 & 0.143 & 0.267 \\ 
   & & 30  & {\bf 0.077} & 0.078 & 0.086 & 0.122 & 0.270 \\ 
   & \multirow{2}{*}{121} & 15 & {\bf 0.078} & 0.080 & 0.098 & 0.121 & 0.258 \\ 
   & & 30 & {\bf 0.069} & 0.071 & 0.089 & 0.105 & 0.264 \\ 
    \multirow{4}{*}{SR} & \multirow{2}{*}{49} & 15 & {\bf 0.099} & 0.099 & 0.099 & 0.159 & 0.297 \\ 
   & & 30  & {\bf 0.094} & 0.095 & 0.094 & 0.138 & 0.290 \\ 
   & \multirow{2}{*}{121} & 15 & {\bf 0.099} & 0.099 & 0.100 & 0.138 & 0.270 \\ 
   & & 30 & 0.095 & {\bf 0.095} & 0.095 & 0.124 & 0.275 \\ 
  \bottomrule
\end{tabular}
\label{Table:sim-MAD}
\end{center}
\end{table}

\begin{table}[htbp]
\caption{The averaged CP and AL over 100 Monte Carlo simulations.}
\begin{center}
\begin{tabular}{ccc|ccccc}
  \toprule
\multicolumn{8}{c}{TR}\\
  \midrule
$n$ & $T$ &  & GLS & GS & TGL & AR & IND \\ 
  \midrule
     \multirow{4}{*}{49} & \multirow{2}{*}{15} & CP & 0.922 & 0.919 & 0.878 & 0.953 & 0.543 \\ 
   & & AL & 0.409 & 0.402 & 0.377 & 0.725 & 0.509 \\ 
   & \multirow{2}{*}{30} & CP &  0.888 & 0.883 & 0.847 & 0.966 & 0.547 \\ 
   & & AL & 0.321 & 0.317 & 0.310 & 0.666 & 0.534 \\ 
     \multirow{4}{*}{121} & \multirow{2}{*}{15} & CP &  0.949 & 0.952 & 0.864 & 0.969 & 0.436 \\ 
   & & AL & 0.398 & 0.410 & 0.376 & 0.663 & 0.364 \\ 
    & \multirow{2}{*}{30} & CP & 0.919 & 0.934 & 0.828 & 0.971 & 0.426 \\ 
   & & AL & 0.315 & 0.343 & 0.305 & 0.597 & 0.365 \\ 
\midrule
\midrule
\multicolumn{8}{c}{SR}\\
  \midrule
$n$ & $T$ &  & GLS & GS & TGL & AR & IND \\ 
  \midrule
     \multirow{4}{*}{49} & \multirow{2}{*}{15} & CP & 0.879 & 0.852 & 0.884 & 0.944 & 0.567 \\ 
   & & AL  & 0.388 & 0.362 & 0.398 & 0.764 & 0.645 \\ 
   & \multirow{2}{*}{30} & CP & 0.770 & 0.730 & 0.771 & 0.958 & 0.555 \\ 
   & & AL & 0.286 & 0.263 & 0.288 & 0.722 & 0.600 \\ 
     \multirow{4}{*}{121} & \multirow{2}{*}{15} & CP & 0.850 & 0.842 & 0.871 & 0.960 & 0.425 \\ 
   & & AL& 0.359 & 0.352 & 0.386 & 0.715 & 0.379 \\ 
    & \multirow{2}{*}{30} & CP & 0.748 & 0.764 & 0.770 & 0.960 & 0.422 \\ 
   & & AL & 0.276 & 0.284 & 0.289 & 0.653 & 0.380 \\ 
  \bottomrule
\end{tabular}
\label{Table:sim-appendix-CPAL}
\end{center}
\end{table}

\end{document}